\documentclass[aps,prstab,preprint,groupedaddress,superscriptaddress,showpacs]{revtex4-1}

\usepackage[USenglish]{babel}			 
\usepackage[pdftex]{graphicx}
\usepackage{ragged2e}
\usepackage{scrextend}
\usepackage{booktabs}
\usepackage{tabularx,setspace}
\newcolumntype{C}{>{\centering\arraybackslash}X}
\usepackage{array,booktabs,amssymb,amsmath,hyperref}
\usepackage{xcolor, soul, colortbl}

\begin{document}


\title{Design of beam optics for the Future Circular Collider $e^+e^-$-collider rings}


\def\KEK{\affiliation{KEK, Oho, Tsukuba, Ibaraki 305-0801, Japan}}
\def\CERN{\affiliation{CERN, CH-1211 Geneva 23, Switzerland}}
\def\BINP{\affiliation{BINP SB RAS, Novosibirsk 630090, Russia}}
\def\UoG{\affiliation{DPNC/Geneva University, CH-1211 Geneva 4, Switzerland}}
\def\SLAC{\affiliation{SLAC, Menlo Park, California 94025, U. S. A.}}

\author{K. Oide}\email[]{email: Katsunobu.Oide@kek.jp}\KEK\author{M. Aiba}\affiliation{PSI 5232, Villigen PSI, Switzerland}\author{S. Aumon}\CERN\author{M. Benedikt}\CERN\author{A. Blondel}\UoG\author{A. Bogomyagkov}\BINP\author{M. Boscolo}\affiliation{INFN/LNF, 00044 Frascati, Rome, Italy}\author{H. Burkhardt}\CERN\author{Y. Cai}\SLAC\author{A. Doblhammer}\CERN\author{B.~Haerer}\CERN\author{B. Holzer}\CERN\author{J.M. Jowett}\CERN\author{I. Koop}\BINP\author{M. Koratzinos}\UoG\author{E. Levichev}\BINP\author{L. Medina}\CERN\author{K. Ohmi}\KEK\author{Y. Papaphilippou}\CERN\author{P.~Piminov}\BINP\author{D. Shatilov}\BINP\author{S. Sinyatkin}\author{M.~Sullivan}\SLAC\author{J. Wenninger}\CERN\author{U. Wienands}\affiliation{ANL, Argonne, Illinois 60439, U. S. A.}\author{D. Zhou}\KEK\author{F. Zimmermann}\CERN


\date{\today}

\begin{abstract}
A beam optics scheme has been designed for the Future Circular Collider-$e^+e^-$ (FCC-ee). The main characteristics of the design are:  beam energy  45 to 175~GeV, 100~km circumference with two interaction points (IPs) per ring, horizontal crossing angle of 30~mrad at the IP and the crab-waist scheme~\cite{ref:cw} with local chromaticity correction. The crab-waist scheme is implemented within the local chromaticity correction system without additional sextupoles, by reducing the strength of one of the two sextupoles for vertical chromatic correction at each side of the IP. So-called ``tapering" of the magnets is applied, which scales all fields of the magnets according to the local beam energy to compensate for the effect of synchrotron radiation (SR) loss along the ring. An asymmetric layout near the interaction region reduces the critical energy of SR photons on the incoming side of the IP to values below~100 keV, while matching the geometry to the beam line of the FCC proton collider (FCC-hh)~\cite{ref:hh} as closely as possible. Sufficient transverse/longitudinal dynamic aperture (DA) has been obtained, including major dynamical effects, to assure an adequate beam lifetime in the presence of beamstrahlung and top-up injection. In particular, a momentum acceptance larger than $\pm2\%$ has been obtained, which is better than the momentum acceptance of typical collider rings by about a factor of 2. The effects of the detector solenoids including their compensation elements are taken into account as well as synchrotron radiation in all magnets. 

The optics presented in this paper is a step toward a full conceptual design for the collider. A number of issues have been identified for further study.
\end{abstract}

\pacs{29.,29.20.db,29.27Bd,}

\maketitle

\section{Requirements and parameters}
The FCC-ee is a double-ring collider to be installed in a common tunnel of $\sim$100~km circumference, as a potential first step before the FCC-hh hadron collider. The beam energy covers a range extending at least from the $Z$-pole (45.6~GeV/beam) to the $t\overline t$ production threshold (175~GeV/beam). The design limits the total SR power at 100~MW, 50~MW for each beam, thus the stored current per beam varies from 1.45~A at the $Z$ to 6.6~mA at the $t\overline t$. This design assumes that the magnet strengths simply scale with the energy, except for the detector solenoid, which will be kept constant at all energies together with the compensation solenoids. A full horizontal crossing angle of $2\theta_x=30$~mrad, together with a crab-waist scheme, is implemented at each IP for all energies, as proposed in Ref.~\cite{ref:lum}. The crossing angle of 30~mrad is sufficiently large to separates the beams and to provide the necessary condition for the crab-waist scheme.  The  critical energy of photons on the incoming side of the IP has been set to below 100~keV at $t\overline t$, from the dipoles upstream up to $\sim$500~m from the IP, and the radiation power from these magnets is about 1~kW. The study of how much radiation can be tolerated by the particle detectors has not yet finished, therefore at this stage we assume similar levels as what was experienced at LEP2, where the critical energy toward the IP was $\sim$80~keV from the last dipole~\cite{ref:LEP2SR}.

This FCC-ee collider ring must have enough DA to store the colliding beam, whose energy spread is drastically increased by beamstrahlung~\cite{ref:fzko,ref:fzmav}, and to maintain the beam current considering the beam lifetime and the ability of the top-up injection scheme~\cite{ref:aiba}. In particular, the dynamic momentum acceptance must be larger than $\pm2$\% at $t\overline t$ to guarantee a sufficiently long beam lifetime in the presence of beamstrahlung~\cite{ref:VT}. A similar momentum acceptance would be required at lower energies, if top-up injection in longitudinal phase space is needed.

\begin{table}[h!]
\begin{minipage}{\textwidth}
   \centering
  \caption{ Machine parameters of FCC-ee. The beam optics simply scales with beam energy. The values in parentheses  correspond to optional cases at each energy. The bunch lengths, the synchrotron tunes, and the RF bucket height shown here are examples corresponding to the RF voltages in the footnotes, and subject to further luminosity optimization.}
   \begin{tabularx}{\textwidth}{|l|C|C|}
  \hline\hline
Circumference [km] & \multicolumn{2}{c|}{99.984} \\
Bending radius of arc dipole [km] & \multicolumn{2}{c|}{11.190} \\
Number of IPs / ring & \multicolumn{2}{c|}{2}\\
Crossing angle at IP [mrad] &\multicolumn{2}{c|}{30}\\
Solenoid field at IP [T]&\multicolumn{2}{c|}{$\pm2$}\\
$\ell^*$ [m] &\multicolumn{2}{c|}{2.2}\\
Local chrom. correction &\multicolumn{2}{c|}{$y$-plane with crab-sextupole effect}\\
Arc cell &\multicolumn{2}{c|}{FODO, $90^\circ/90^\circ$}\\
Momentum compaction $\alpha_p$ [$10^{-6}$]&\multicolumn{2}{c|}{6.99}\\
$\beta$-tron tunes $\nu_x/\nu_y$&\multicolumn{2}{c|}{$387.08/387.14$}\\
Arc sextupoles &\multicolumn{2}{c|}{292 pairs per half ring}\\
RF frequency [MHz] &\multicolumn{2}{c|}{400}\\
\hline
Beam energy [GeV] & 45.6 & 175\\
\hline
SR energy loss/turn [GeV] & 0.0346 & 7.47\\
Longitudinal damping time [ms] & 440 & 8.0\\
Polarization time [s] &  $9.2\times10^5$ &1080 \\
Current/beam [mA] & 1450 & 6.6\\
Bunches/ring & 30180 (91500) & 81 \\
Minimum bunch separation [ns] & 10 (2.5) & 2000 \\
Total SR power [MW] & 100.3 & 98.6\\
Horizontal emittance $\varepsilon_x$ [nm] & 0.86 & 1.26\\
$\varepsilon_y/\varepsilon_x$ with beam-beam [\%] & 0.6 & 0.2 \\
$\beta^*_x$ [m] & 0.5 (1) & 1 (0.5)\\
$\beta^*_y$ [mm] & 1 (2) & 2 (1)\\
Energy spread by SR [\%] & 0.038 & 0.141\\
Bunch length by SR [mm] & 2.6\footnote{\label{fn1}for RF voltage $V_c=88$ MV}& 2.4\footnote{\label{fn2}for RF voltage $V_c=9.04$ GV}\\
Synchrotron tune $\nu_z$ & $-0.0163$\footref{fn1} & $-0.0657$\footref{fn2}\\
RF bucket height [\%] & 2.3\footref{fn1} & 11.6\footref{fn2} \\
Luminosity/IP [$10^{34}/{\rm cm}^2{\rm s}$] &  210 (90) & 1.3 (1.5) \\
\hline\hline
   \end{tabularx}
   \label{params}
\end{minipage}
\end{table}

 Table~\ref{params} shows the machine parameters. The optics simply scales with the energy. The minimum achievable $\beta^*_{x,y}$ are (0.5~m, 1~mm) at all energies, and the values should be chosen to maximize the luminosity performance at each energy. Actually the luminosity gain of (0.5~m, 1~mm) compared to (1~m, 2~mm) is estimated to be small at $t\overline t$ as shown in Table~\ref{params}, and $\beta^*_{x,y}=$~(1~m, 2~mm) is considered to be the baseline at $t\overline t$.
 
\section{Layout}
\begin{figure}[h!]
   \centering
   \includegraphics[width=350pt]{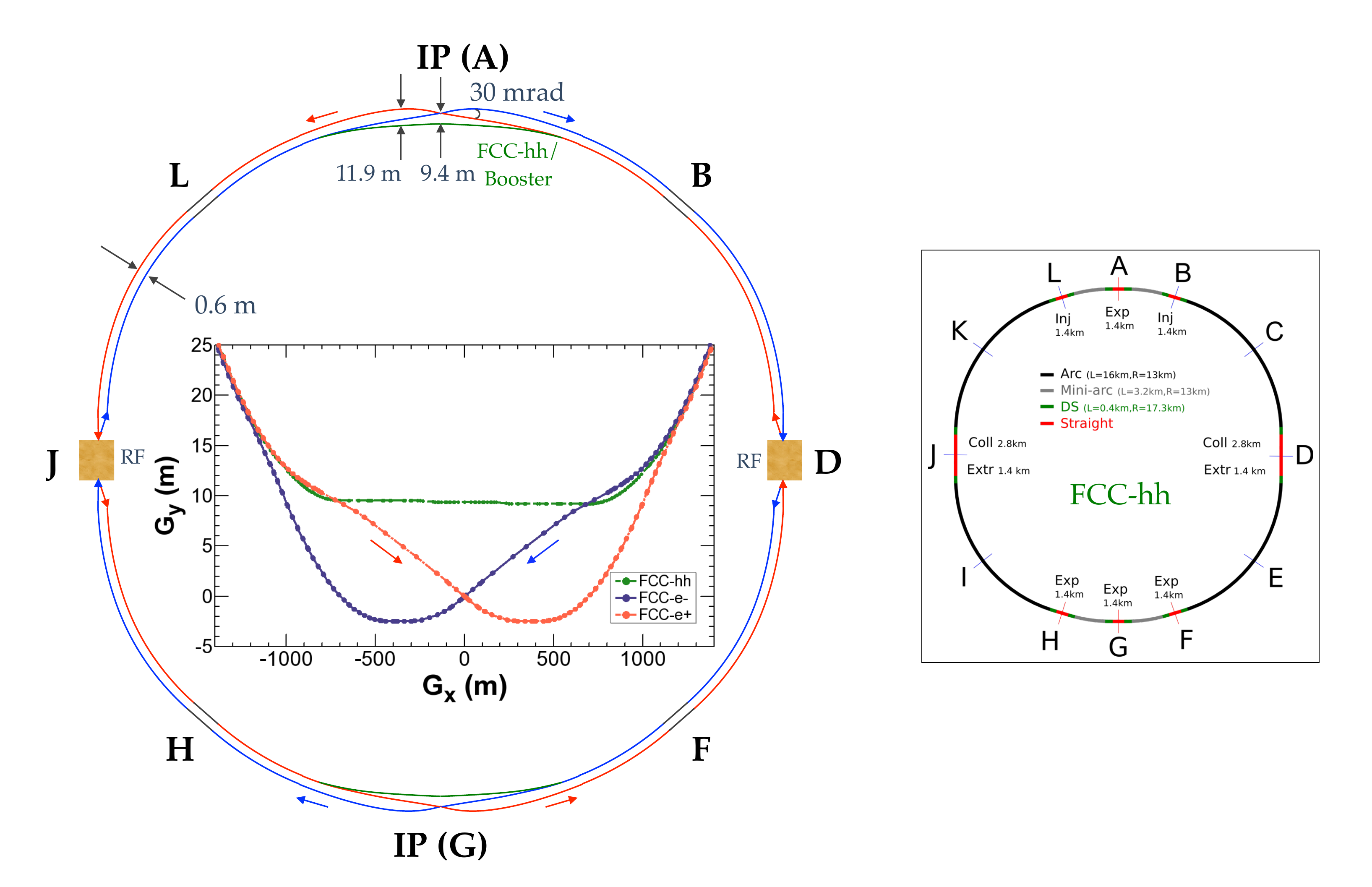}
   \caption{The left figure shows the schematic layout of the FCC-ee collider rings. The two rings are horizontally separated by 0.6 m in the arc, and their center is placed on the center of the FCC-hh hadron rings. The straight sections correspond to the hadron ring, shown in the right figure. Two IPs are located in the straight sections A and G, and the RF sections are located in D and J. There exist intermediate straight sections B, F, H, L  in the arcs. Beams cross over in the RF sections. The south IP (G) is shown enlarged in the middle of the left figure. The green line indicates the FCC-hh beam line, along with also the $e^+/e^-$ booster synchrotron for FCC-ee can be placed.    \label{fig:Layout}}
\end{figure}

The schematic layout of the FCC-ee rings is shown in Fig.~\ref{fig:Layout}. The basic geometry just follows the current layout of the FCC-hh ring~\cite{ref:shulte}. The $e^+e^-$ rings are placed side by side. In the arc sections, the center of the $e^+e^-$ rings is exactly placed on the center of the hh-rings, while these are offset by about 1~m in the straight sections, except for the interaction region (IR). The layout in the IR is greatly constrained by the requirement on the incoming synchrotron radiation. To implement a crossing angle at the IP, the beam must come from the inner ring to the IP, then be bent strongly after the IP to merge back close to the opposing ring. Thus the IP of the $e^+e^-$ rings is displaced towards the outside relative to the hh-beam. The magnitude of the displacement of the IP depends on the limit of the critical energy of photons hitting the IP. The design shown in Fig.~\ref{fig:Layout} has a displacement of  9.4~m. Therefore in th IR, $e^+e^-$ beams separate from the hh-beam line reaching a maximum deviation of 11.9~m at around $\pm$400~m from the IP. The outgoing beam returns to the hh-beam line at around $\pm1.2$~km from the IP. For this region a wide tunnel or double tunnels are needed. These separation distances can be reduced if the criteria for the SR are relaxed, as shown in an alternative design~\cite{Anton}. The shift of the IP allows installing the booster synchrotron along the hh-beam line so as to bypass the $e^+e^-$ detector.

At each IP, the beam must come from the inside. Thus the beams must cross over somewhere between the IPs. This is naturally done if we have the RF cavities common to both beams, which is especially beneficial in the case of $t\overline t$. If only a half of each ring is filled by bunches, each beam passes the cavities without seeing the other beam. As the number of bunches per ring are 80 for the $t\overline t$, much less than the number of RF buckets ($\sim$133,400), such a filling scheme is not an issue. Moreover, the RF cavities are loaded with the two beams in succession, keeping the transient loading low. At lower energies, such a common RF is not necessary, and a simple crossing without interaction can be implemented.

\section{Beam Optics}
\subsection{Arc}
\begin{figure}[h!]
   \centering
   \includegraphics[width=250pt]{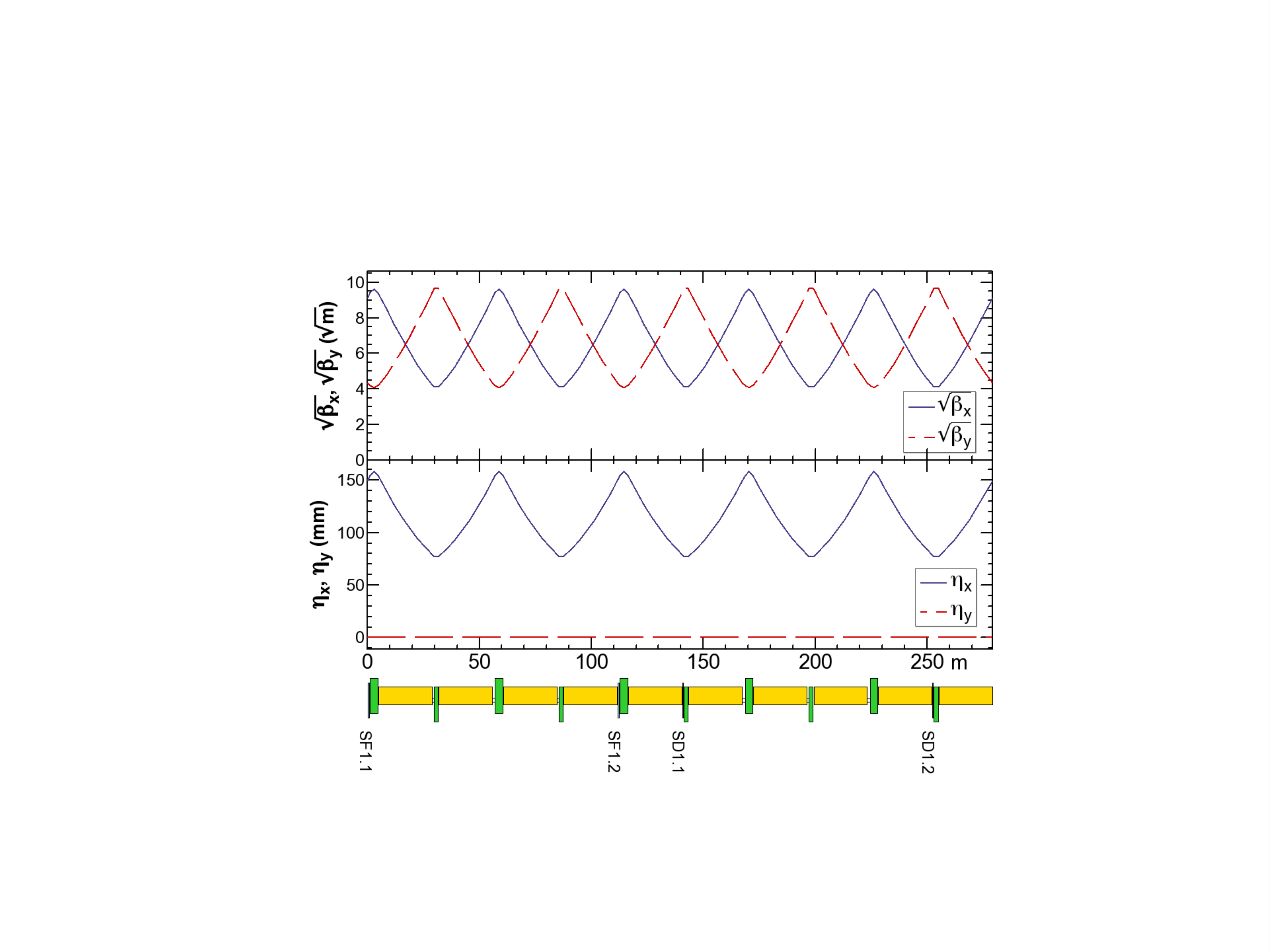}
   \caption{The beam optics of the arc unit cell of FCC-ee. Upper and lower plots show $\sqrt{\beta_{x,y}}$ and dispersions, respectively. Here we show a supercell  consisting of 5 $90^\circ/90^\circ$ FODO cells. The horizontal/vertical  sextupoles SF1.1/SD1.1 are paired with SF1.2/SD1.2 via $-I$ transformations.
   \label{ArcCell}}
\end{figure}

The arc optics consists of $90^\circ/90^\circ$ FODO cells as shown in Fig. \ref{ArcCell}. Two non-interleaved families of sextupole pairs, with a $-I$ transformation between sextupoles, are placed in a supercell consisting of 5 FODO cells.  This scheme has been applied successfully at B-factories for more than 15 years~\cite{KEKB}. The number of cells is determined by the equilibrium horizontal emittance, resulting in 292 independent sextupole pairs per half ring. So far we have assumed a complete period 2 periodicity of the ring optics. The length of quadrupoles must be chosen by considering the power consumption as well as the effect of the synchrotron radiation on the DA, which is discussed in Appendix~\ref{app:aSRQ}. As the non-interleaved sextupole scheme cancels the primary transverse nonlinearity from the sextupoles~\cite{KLB}, the resulting on-momentum horizontal dynamic aperture reaches 70$\sigma_x$ at the $Z$ energy as shown in Fig. \ref{DA}(b). At the $t\overline t$ energy, however, the peak disappears due to the SR loss in the quadrupoles (Fig. \ref{DA}(a)).

There is a possibility to use a combined function dipole magnets for the arc cell to reduce the number of cells in the arc, while maintaining the same horizontal emittance. A study described in Appendix~\ref{app:CD} demonstrates a reduction in the number of cells by about 30\% for a field gradient in the dipoles that changes the longitudinal damping partition from 2.0 to 0.6. 

\subsection{IR}
One of the challenges of beam optics for the FCC-ee collider is providing the dynamic aperture with small $\beta$-functions at the IP down to $\beta^*_{x,y} =$ (0.5~m, 1~mm). Although these values are still higher than those in modern B-factories~\cite{SuperKEKB}, the associated vertical chromaticity around the IP is comparable, since the distance, $\ell^*$, from the face of the final quadrupole magnet to the IP is much longer than those in B-factories. Also especially at the $t\overline t$ energy, the beamstrahlung caused by the collisions requires a very wide momentum acceptance of $\pm2\%$. The transverse on-momentum  dynamic aperture must be larger than $\sim15\sigma_x$ to enable top-up injection in the horizontal plane.

\begin{figure}[h!]
   \centering
   \includegraphics*[width=300pt]{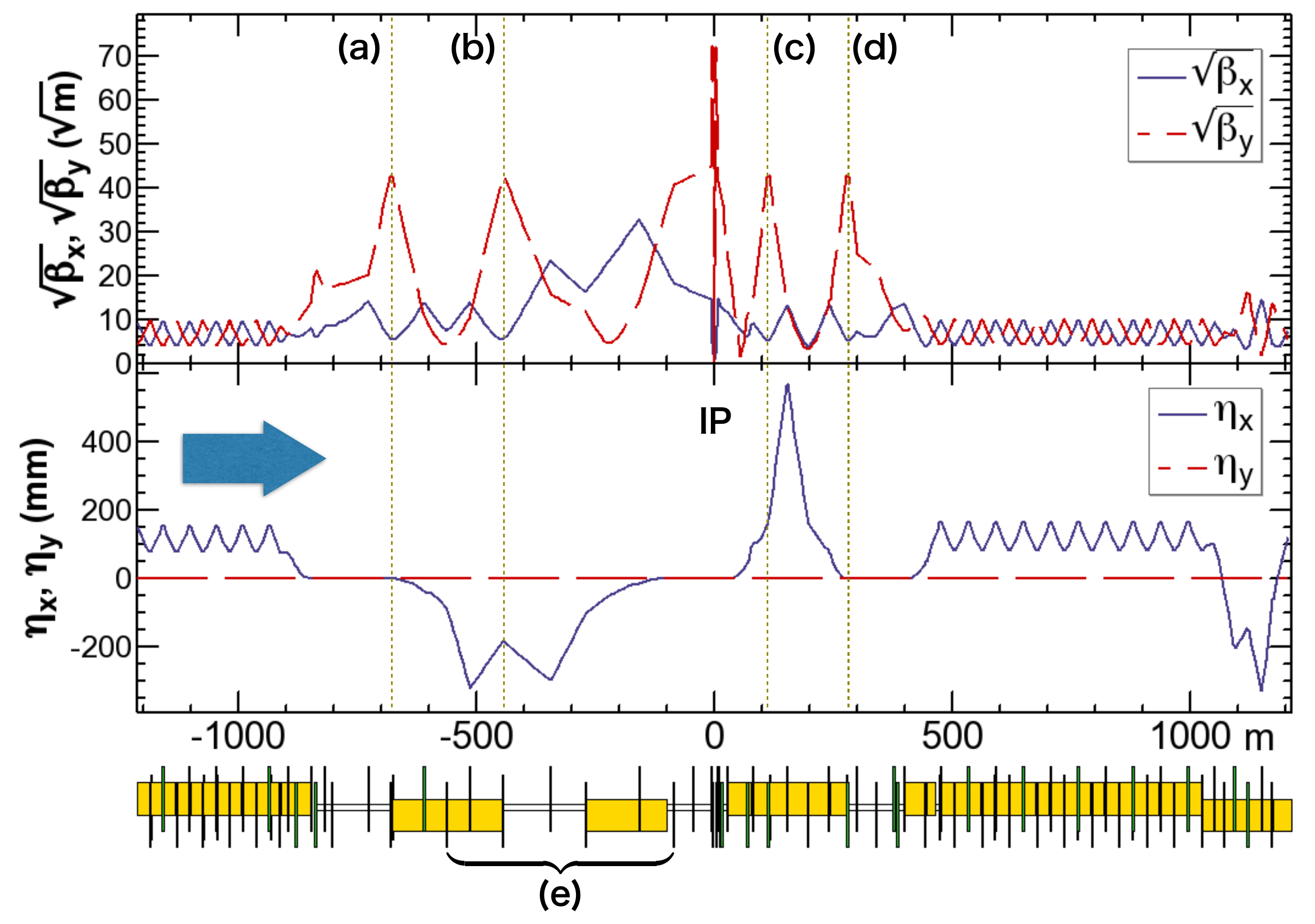}
   \caption{ The beam optics of the FCC-ee IR, corresponding to $\beta^*_{x,y} =$ (1~m, 2~mm).  Upper and lower plots show $\sqrt{\beta_{x,y}}$ and dispersions, respectively. The beam passes from the left to the right in this figure. The optics is asymmetric to suppress the synchrotron radiation toward the IP. Dipoles are indicated by yellow boxes, and those in region (e) have a critical energy of the SR photon below 100~keV at the $t\overline{t}$.  Sextupoles for the LCCS are located at (a--d), and sextupoles at (a,d) play the role of crab sextupoles. \label{IR}}
\end{figure}

Figure~\ref{IR} shows the optics of the IR for $\beta^*_{x,y} =$ (1~m, 2~mm). It has a local chromaticity correction system (LCCS) only in the vertical plane at each side of the IP. The sextupole magnets are paired at each side, and only the inner ones at (b,c) have nonzero horizontal dispersion~\cite{oddD}. The outer ones at (a,d) do not only cancel the geometrical nonlinearity of the inner ones, but also generate the crab waist at the IP by choosing their phase advance from the IP as $\varDelta\psi_{x,y}=(2\pi, 2.5\pi)$, as described in Appendix~\ref{app:crab}. The incorporation of the crab sextupoles into the LCCS saves space and reduces the number of optical components.

\begin{figure}[h!]
   \centering
   \includegraphics*[width=300pt]{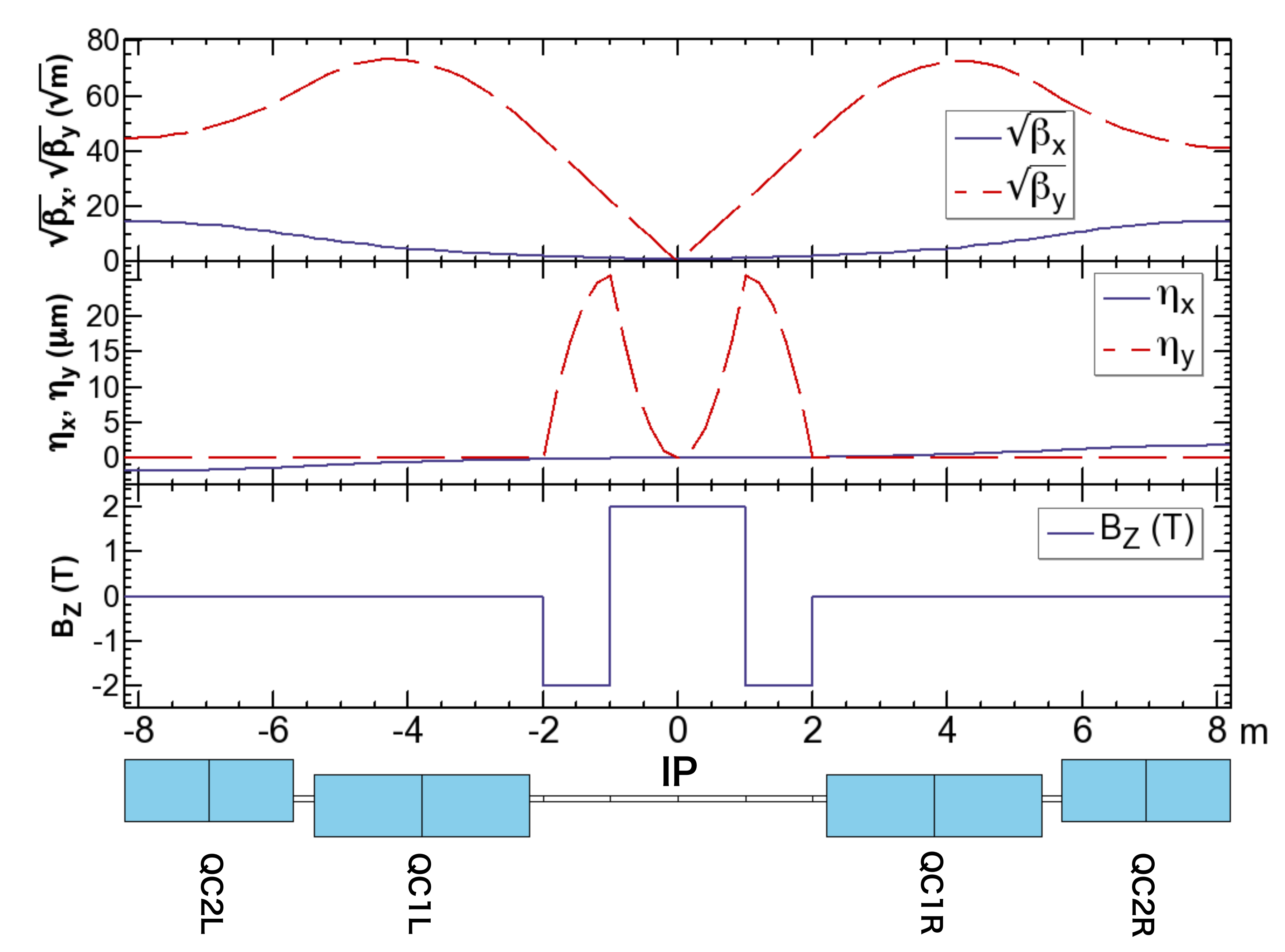}
   \caption{ The beam optics of FCC-ee around the IP . Plots show $\sqrt{\beta_{x,y}}$ (top), horizontal/vertical dispersions (middle), and the solenoid field (bottom), respectively. QC1/QC2 denote the vertical/horizontal focusing quadrupoles. The distance $\ell^*$ between the face of QC1 and the IP is 2.2 m. The vertical dispersion is locally confined.\label{IP}}
\end{figure}

The beam lines at the interaction region are separate for the two beams. There are no common quadrupoles in the IR. So far $\ell^*$ is chosen to be 2.2~m, which is sufficient for two independent final quadrupoles with a 30~mrad crossing angle. Also $\ell^*=2.2$~m has been accepted by the physics detector group as a working assumption~\cite{ref:azzi}. This is the subject of further study depending on the detailed design of the detector and its interface with the machine. The solenoids are common for two beams, and they are locally compensated with counter solenoids to cancel $\int B_zdz$ between the IP and each face of the final quadrupole, as shown in Fig.~\ref{IP}. No vertical orbit, vertical dispersion, and $x$-$y$ couplings leak to the outside for any particle at any energy. So far we have assumed such a perfect compensation. The final quadrupoles have a field gradient of 100~T/m, and the detector field at the locations of the quadrupoles is canceled by additional shielding solenoids, which completely remove the longitudinal field on the quadrupoles. We assume a step-function profile of the solenoid field as shown in Fig.~\ref{IP}. A complete conceptual analysis of all magnetic elements around the IP has be performed~\cite{Koratzinos}, which verifies the effect on the optics is minimal. The vertical emittance is increased due to the fringe field of the compensating solenoid in combination with the horizontal crossing angle. The increase becomes largest at the $Z$ energy as we assume that the solenoid field is independent of the beam energy. The increase of the vertical emittance is below 0.2~pm, for 2 IPs, with the step-function profile assuming 10~cm for the length of fringe. The realistic component analysis also gives comfortably smaller value of than the design emittance at collision.

The critical energy of SR photons from the dipoles up to 500~m upstream of the IP is set below 100~keV at $t\overline t$. There is no dipole magnet before the IP up to 100~m upstream.

\subsection{RF section and other straight sections}
\begin{figure}[h!]
   \centering
   \includegraphics*[width=300pt]{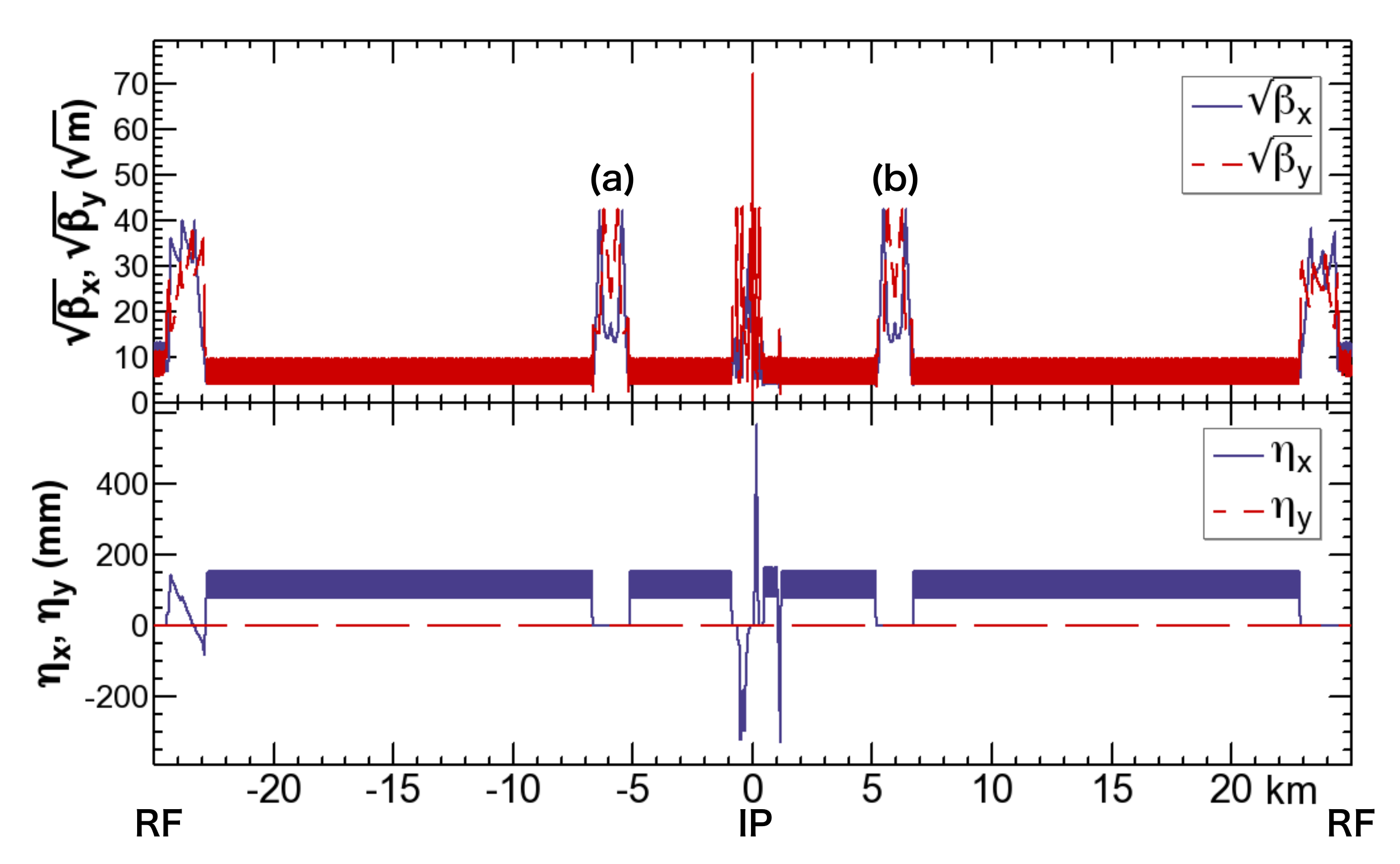}
   \caption{ The beam optics of the FCC-ee half ring, corresponding to $\beta^*_{x,y} =$ (1~m, 2~mm). Upper/lower plots show $\sqrt{\beta_{x,y}}$ and horizontal/vertical dispersions, respectively. These plots start and end in the middle of the RF sections, and the IP is located at the center. Sections marked by (a,b) correspond to the intermediate straight sections B, F, H, L in Fig.~\ref{fig:Layout}.\label{fig:Ring}}
\end{figure}

Figure~\ref{fig:Ring} presents the beam optics for the half ring. The RF sections are located in the straight sections J and D in Fig.~\ref{fig:Layout}. At $t\overline t$, an acceleration voltage of $\sim$4.5~GV is needed, so the length of the RF section would be about 1~km. Both beams pass through a common RF section. A combination of electrostatic separator and a dipole magnet deflects only the outgoing beam so as to avoid SR toward the RF cavities. The quadrupoles within the RF section are common to both beams, but are still compatible with the overall tapering scheme, if their strengths are chosen symmetrically.

The usage of the intermediate straight sections in the middle of the arc has not been determined. Some of them can be used for injection, dump, collimation, etc. The current optics for them have not been finalized.

\section{Dynamic Aperture\label{DAsec}}
The dynamic aperture (DA) for the optics described in this paper has been estimated using the computer code SAD~\cite{SAD}, considering a number of effects listed in Table~\ref{effects}. Among them, synchrotron radiation plays an essential role. While the radiation loss in dipoles improves the aperture, especially at $t\overline t$, due to the strong damping, the radiation loss in the quadrupoles for particles with large betatron amplitudes reduces the dynamic aperture. This is due to the induced synchrotron motion through the radiation loss as described in Appendix~\ref{app:aSRQ}. This effect is mostly noticeable in the horizontal arc quadrupoles. Thus the length of the arc quadrupoles must be sufficiently long. The final focus quadrupole has the similar effect on the vertical motion in the case of $\beta^*_y=1$~mm due to the large $\beta_y$ and the strong field gradient in the quadrupole.

\definecolor{light-gray}{gray}{0.9}
\definecolor{LightCyan}{rgb}{0.88,1,1}
\definecolor{White}{rgb}{1,1,1}


\begin{table}[htb!]
   \centering
\begin{minipage}{\textwidth}
  \caption{Effects taken into account in the estimation of the dynamic aperture.   \label{effects}}
 \renewcommand{\arraystretch}{1.2}
       \setstretch{0.9}
\renewcommand{\tabularxcolumn}[1]{>{\small}m{#1}}

\begin{tabularx}{\textwidth}{|>{\raggedright\arraybackslash}X|c|>{\raggedright\arraybackslash}X|}
\rowcolor{White}
\hline\hline
Effect & Included? & Significance at $t\overline t$\\
  \hline
Synchrotron motion & Yes & {\bf Essential} \\ \rowcolor{LightCyan}
Radiation loss in dipoles& Yes & {\bf Essential} -- improves the aperture\\
Radiation loss in quadrupoles \footnote{See Appendix \ref{app:aSRQ}} & Yes &  {\bf Essential} -- reduces the aperture \\ \rowcolor{LightCyan}
Radiation fluctuation \footnote{not included in the optimization, see Appendix \ref{fluct}} & Yes &  {\bf Essential}\\
Tapering & Yes &  {\bf Essential} \\ \rowcolor{LightCyan}
Crab waist & Yes & transverse aperture is reduced by $\sim20\%$\\
Solenoids & Yes & minimal, if locally compensated\\ \rowcolor{LightCyan}
Maxwellian fringes~\cite{ref:Forest} & Yes & small\\
Kinematical terms & Yes & small\\ \rowcolor{LightCyan}
Higher order fields/errors/misalignments & No & {\bf Essential}, development of correction/tuning scheme is necessary\\ 
\hline\hline
   \end{tabularx}
\end{minipage}
\end{table}

The DA has been optimized by going through the settings of sextupoles using particle tracking with a downhill simplex method scripted within SAD. All effects in Table~\ref{effects} are included in the optimization, except for radiation fluctuation, which requires a large number of samples of random numbers and cannot be taken into account with the available computing resources. The effect of radiation fluctuation is evaluated separately after the optimization, as discussed in Appendix~\ref{fluct}. Figure~\ref{DA} shows a result of such an optimization. The goal of the optimization is to maximize the figure of merit $F$ as defined below at each beam energy in the $z$-$x$ plane. The results are shown in Fig.~\ref{DA}(a) and (b) for ${t\overline t}$ and $Z$ energies. The transverse apertures, Fig. \ref{DA}(c) and (d) are just the result of the optimization in the $z$-$x$ plane. 

The DA shown here is obtained from the survival of particles in a set of initial conditions with a finite amplitude starting at the middle of the RF section, which is the middle point between IPs. A set of initial coordinates are chosen in the $z$-$x$ plane as
\begin{eqnarray}
z_{1ik}&=&\left(i/n_x A_x \sigma_x,0,i/n_x A_x \sigma_y,0,A_z \sigma_\delta\cos(k \pi/2n_z)\right)\ ,\label{inix}\\
z_{2ik}&=&\left(0,i/n_x A_x \sigma_{p_x},0,i/n_x A_x \sigma_{p_y},0,A_z \sigma_\delta\cos(k \pi/2n_z)\right)\ ,\label{inipx}
\end{eqnarray}
where $i$ and $k$ are integers running from $-n_{x,z}$ through $n_{x,z}$, respectively, and the coefficients $A_{x,z}$ are the maximum amplitudes to investigate. The two sets $z_{1ik}$ and $z_{2ik}$ above correspond to the apertures $\varDelta x$ and $\varDelta p_x$ in Fig.~\ref{DA}(a,b), respectively. The parameters used to produce the results in Fig.~\ref{DA}(a/b) were $n_x=50/50$, $n_z=15/15$, $A_x=25/65$, and $A_z=15/52$, respectively, which gave 6,262 initial condition combinations of $z_{1ik}$ and $z_{2ik}$. The figure of merit $F$ for the optimization is expressed as
\begin{equation}
F=\sum_k \sup\left\{i_2-i_1 | S(z_{1ik})\wedge \forall i\in[i_1,i_2]\right\}+ \sup\left\{i_2-i_1 | S(z_{2ik})\wedge \forall i\in[i_1,i_2]\right\}\ ,
\end{equation}
where $S(z)$ is that the particle starting with the initial condition $z$ survives after the specified number of revolutions. Thus the cosine-distribution in the momentum direction in Eqs.~(\ref{inix},\ref{inipx}) gives more weight towards the extremities of the momentum acceptance.

The resulting DA satisfies the requirements for both beamstrahlung and top-up injection, at least without field errors and misalignments. We have assumed that the booster injects a beam with the same $\varepsilon_x$ as the collider ring and with $\varepsilon_y/\varepsilon_x=10\%$. The optimization must be done for each setting of $\beta^*_{x,y}$, $\beta$-tron tunes, and the beam energy.  The number of initial conditions that can be studied is basically limited by the available computing resources. A larger number is always better, but when we doubled $n_z$ and the number of revolutions from Fig.~\ref{DA}, the change in the resulting DA was tiny. Also a beam-beam simulation including lattice and beamstrahlung indicates that the DA is sufficient to hold the beam with beamstrahlung at $t\overline t$ as discussed below.

\begin{figure}[hbt!]
   \centering
  \includegraphics*[width=450pt]{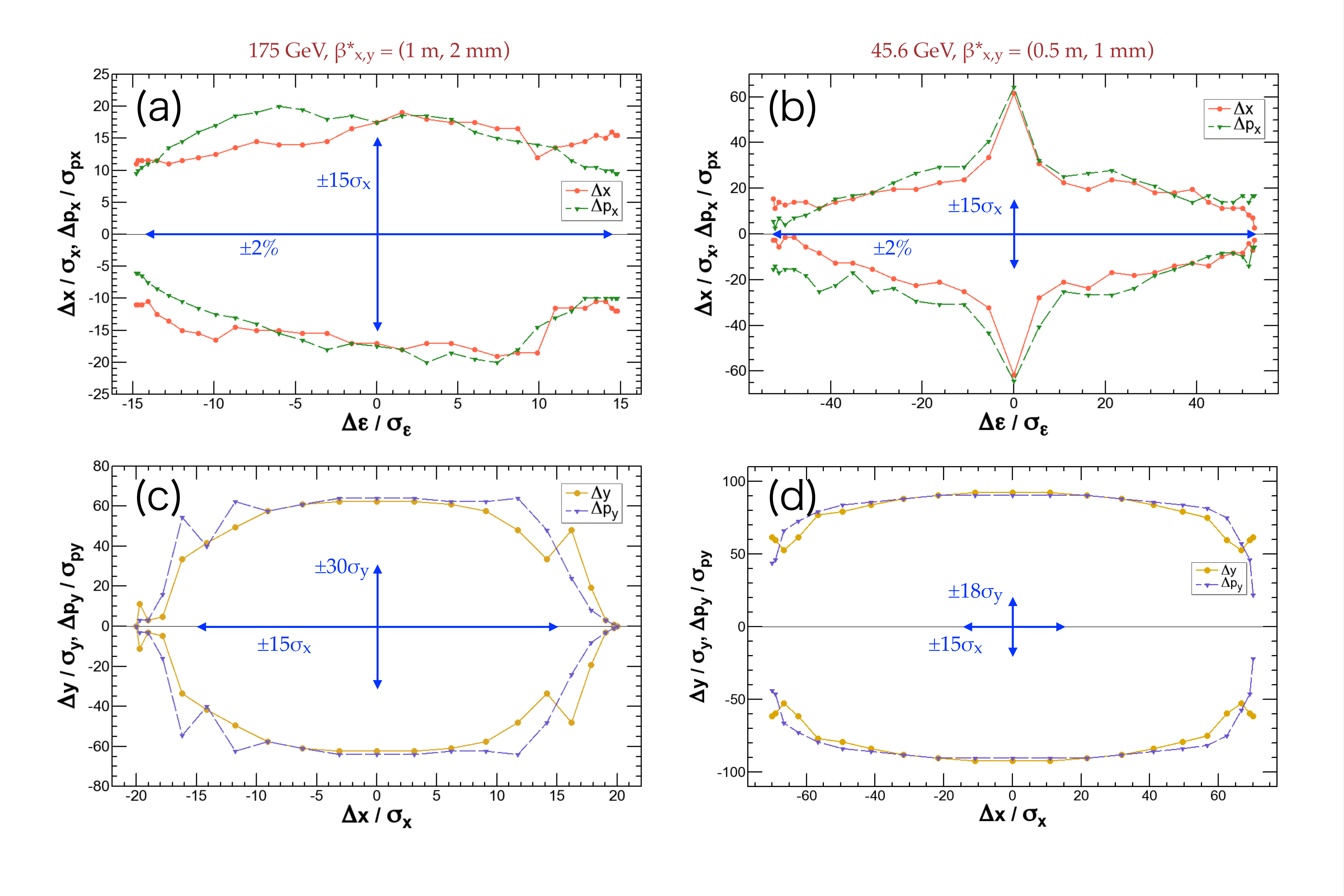}
   \caption{Dynamic apertures after an optimization of sextupoles via particle tracking. (a, c): $\beta^*_{x,y}=$(1 m, 2mm), 50 turns at $t\overline t$, (b, d): $\beta^*_{x,y}=$(0.5 m, 1mm), 2,650 turns at $Z$. (a, b): $z$-$x$ plane with $J_y/J_x=0.2\%$ for (a) and $0.5\%$ for (b). (c, d): $x$-$y$ plane with $\delta=\varDelta\varepsilon=0$. The aperture was searched both in (a,b) $x$ and $p_x$ or (c,d) $y$ and $p_y$ directions. The initial conditions are chosen according to Eqs.~(\ref{inix},\ref{inipx}). The number of turns is chosen to correspond to about 2 longitudinal damping times at each energy. The blue lines show the DAs required for the beamstrahling and the top-up injection. Effects in Table~\ref{effects} are taken into account except for the radiation fluctuation and the beam-beam effect.}
   \label{DA}
\end{figure}

So far all sextupole pairs have been used independently in the optimization, thus the degree of freedom for the optimization is nearly 300. The periodicity of super period 2 of the ring is kept. It has not been verified whether the large number of sextupole families is really necessary. Solutions with fewer families have also been investigated~\cite{Anton,Haerer}.

 One of the goals of this design is to ensure sufficiently large dynamic aperture and momentum acceptance in presence of beamstrahlung. A study was done~\cite{Zhou} using a model of strong-weak beam-beam interaction implemented in SAD, which simulates a realistic photon spectrum of beamstrahlung~\cite{ref:ohmibs}.  The simulation was done at the $t\overline t$ energy including the lattice and all the effects listed in Table~\ref{effects} up to the design beam intensity.  No particle loss was observed with the $\beta^*_{x,y}=$ (1~m, 2~mm) lattice, when tracking 10,000 macro particles for up to 4,000 turns, which corresponds to a beam lifetime $\tau\gtrsim 3$~hours.  This lifetime  is much longer than that given by other processes such as radiative Bhabha scattering, which is $\sim$57~minutes with the luminosity in Table~\ref{params}, using the cross section $\sigma_{ee} \approx 0.15$~b~obtained by BBBREM \cite{ref:bhabha} for 2\% momentum acceptance. In the case of $\beta^*_{x,y}=$ (0.5~m, 1~mm), the lifetime with beamstrahlung becomes 40 minutes, and the total lifetime is still within the capacity of the top-up injection.  Thus the dynamic aperture of the design satisfies the requirement, that is not significantly affected by beamstrahlung at least without any machine errors.

\section{Remaining Issues \label{sect:ri}}
The design of optics presented in this paper is a basic step toward the design of the FCC-ee circular collider. There are a number of  issues remaining to be addressed, such as:
\begin{itemize}
\setlength\itemsep{0em}
\item Development of correction/tuning schemes for the emittance and the dynamic aperture to mitigate the possible higher-order fields, machine errors, and misalignments.
\item Development of practical algorithm for the optimization of a large number of sextupole families, under the presence of machine imperfection and its time variation.
\item Further studies on beam-beam effects with machine imperfections.
\item Refinement of the IR region considering the machine-detector interface.
\item Iterations taking into account progressively more complete hardware designs of the RF, vacuum chambers, magnets, beam instrumentation, etc.
\item Development of the injection scheme and associated optics. See also Ref.~\cite{ref:aiba}.
\item Detailed study of the beam background for the physics detector as well as development of the collimation system with  associated beam optics.
\end{itemize}

\section{Conclusions}
\begin{enumerate}
\item We presented the design for a highest-energy circular $e^+e^-$ collider (FCC-ee) with ultra-low $\beta^*$ of 1 mm and more than $\pm2\%$ dynamic momentum acceptance.
\item This design features a local chromatic correction for the vertical plane.  The dynamic aperture was optimized by varying the strengths of about 300 independent $-I$ sextupole pairs in the arcs.
\item A crab-waist scheme was implemented by reducing the strength of an existing sextupole in the chromatic correction section with proper betatron phases, instead of adding another dedicated sextupoles.
\item Synchrotron radiation is accommodated by tapering the magnet strengths in the arcs, and by a novel asymmetric IR/final-focus layout.
\item The RF system is concentrated in two straight sections. A common system provides maximum voltage for $t\overline t$ running, where operation requires only few bunches. Two separate RF systems, one for either beam, are used at lower beam energies.
\item The optics was designed to match the footprint of a future hadron collider (FCC-hh) along the arcs. Due to the asymmetric IR layout the $e^+e^-$ IP is displaced transversely by about 9~m from the hadron IP. This allows a lepton detector to be installed in the same cavern.
\item The optics, the footprint, and the dynamic aperture are compatible with a top-up injection mode of operation based on a full-energy booster ring installed in the same tunnel and, in the IR, following the path of the hadron collider ring. 
\end{enumerate}

\newpage
\appendix
\section{Lattice with Combined Function Dipole \label{app:CD}}
\begin{figure}[h!]
   \centering
   \includegraphics*[width=300pt]{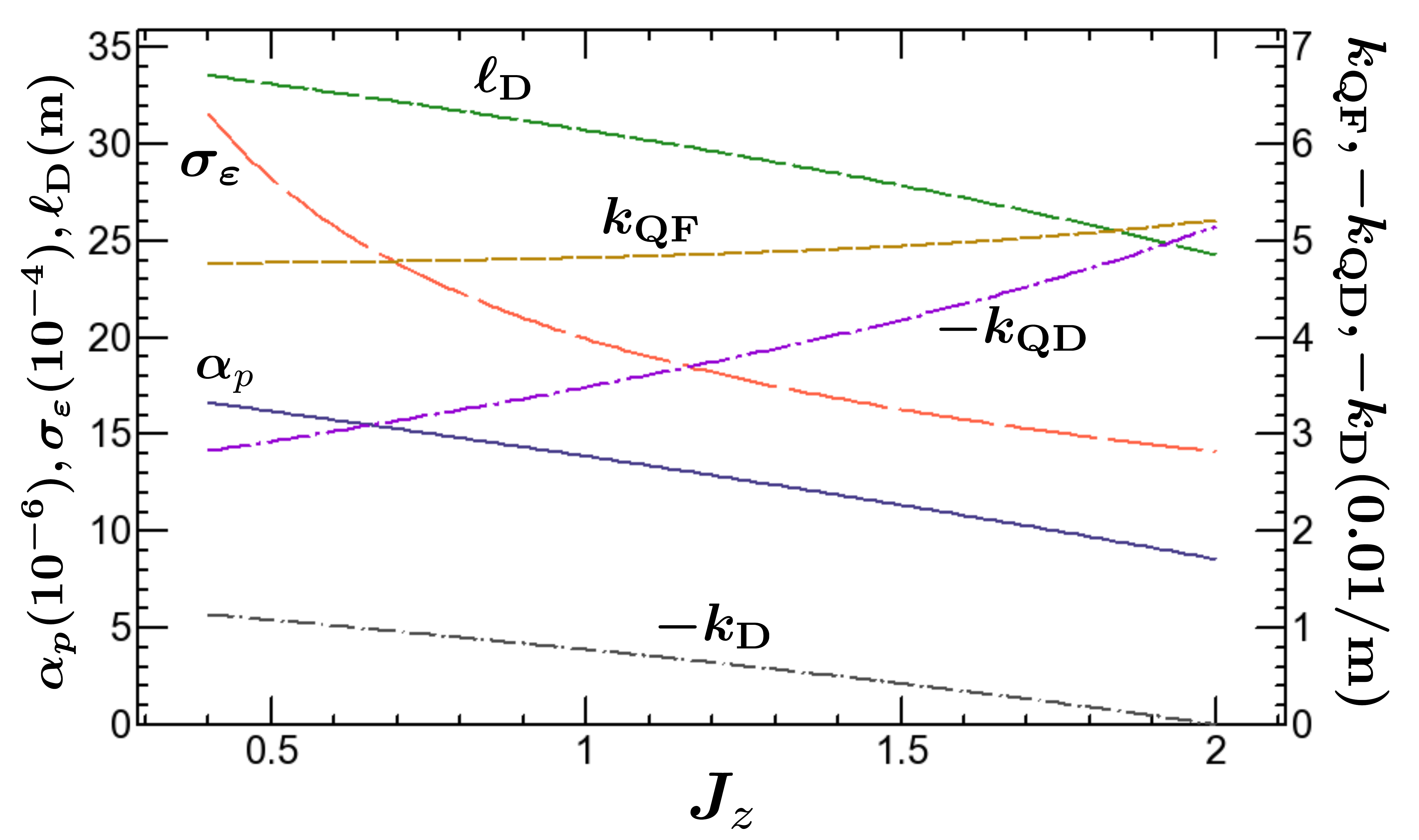}
   \caption{Variation of the momentum compaction factor $\alpha_p$, momentum spread $\sigma_\varepsilon$, length of the dipole magnet $\ell_{\rm D}$, and focusing strengths of dipole, horizontal/vertical focusing quadrupoles $k_{\rm D,QF,QD}$ ($=B'\ell/B\rho$) of $90^\circ/90^\circ$ FODO cell as functions of $J_z$. The horizontal equilibrium emittance is fixed equal to  $\varepsilon_x=1.25$~nm at 175~GeV. 
   \label{combined}}
\end{figure}

 Using combined function dipoles is a well-known method for controling the equilibrium horizontal emittance, momentum compaction factor, etc. by changing the damping partition number. Figure~\ref{combined} plots the variation of several parameters of a unit cell, as functions of the longitudinal damping partition $J_z$ with a fixed horizontal emittance and phase advances. As shown in this figure, reducing $J_z$ from a flat dipole ($J_z=2$) makes $\ell_{\rm D}$ longer, which means a longer cell length and fewer cells in a ring. The associated larger $\alpha_p$ provides a longer bunch length for a given voltage, which is favorable for the beam-beam performance, especially at the $Z$. It also makes $\sigma_\varepsilon$ larger, which relaxes the enlargement ratio due to beamstrahlung. The larger $\sigma_\varepsilon$, however, will reduce the level of  transverse polarization, which is essential for the beam energy calibration at the $Z$ and $WW$ energies.
 
 Table~\ref{CDtab} shows a comparison of the effects on the lattice design between flat and $J_z=0.6$. The resulting design seems to have merits, except for the larger energy spread affecting the beam energy calibration capability. The number of cells, thus of quadrupole is reduced by about 30\%. The total synchrotron radiation loss reduces by 10\%, due to the improved packing factor of dipoles, resulting in 10\% higher luminosity for a give SR power. The larger horizontal dispersion at the sextupoles should improve dynamic aperture. This lattice still maintains a capability for tuning, as two quadrupoles remain in a cell. Further study is needed for the machine tolerance and tuning as well as the technical design of the combined-function dipole magnet.
 
\begin{table}[h!]
   \centering
  \caption{Comparison between a lattice with combined ($J_z=0.6$) and flat ($J_z=2$) dipoles at $t\overline{t}$. The bold faces highlight the merits of the combined dipole scheme. \label{CDtab}}
   \begin{tabular}{|l |c | c|}
  \hline\hline
  $J_z$ & 0.6 & 2\\
  \hline
  Number of FODO cells&{\bf 1062}&1442\\
  Length of dipole $\ell_{\rm D}$ [m]&33.9&23.1\\
  Horizontal dispersion at SF [cm]&{\bf 29.6}&16.3\\
  One turn energy loss $U_0$ [GeV] & {\bf 7.09} & 7.74\\
  Momentum spread $\sigma_\varepsilon$ [\%]& 0.24&0.14\\
  Momentum compaction $\alpha_p$  [$10^{-6}$]&12.8&7.2\\
  Bunch length $\sigma_z$  [mm]&{\bf 5.0}&2.4\\
  RF voltage $V_c$ [GV]&9.6&9.4\\
  synchrotron tune $\nu_z$ &$-0.10$&$-0.068$\\
  \hline\hline
  \end{tabular}
  \end{table}
  
\section{Use of Sextupoles in the Chromatic Correction System for Crab Waist\label{app:crab}}

\begin{figure}[h!]
   \centering
   \includegraphics[width=300pt]{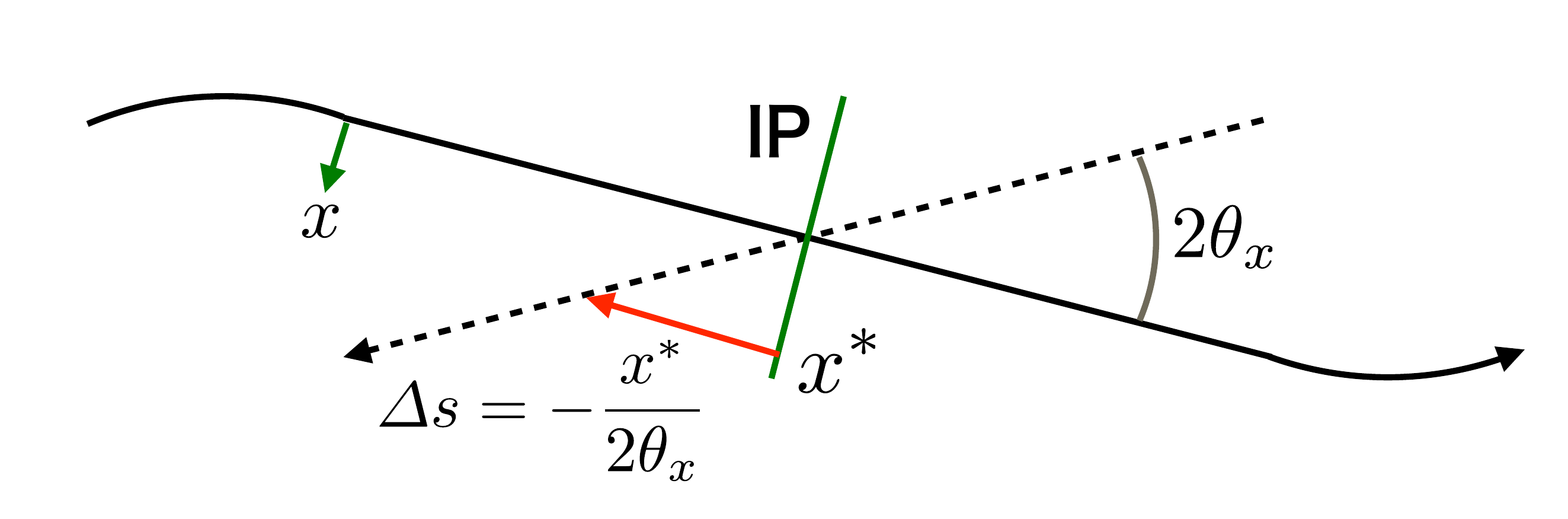}
   \caption{The crab-waist scheme: Two beams cross at the IP with a horizontal crossing angle $2\theta_x$. The crab-waist scheme shifts the vertical waist of a beam (solid) in $s$-direction by $\varDelta s$ onto the center of the other beam (dashed).
   \label{crabsext}}
\end{figure}

The crab-waist scheme shifts the vertical waist of a beam by
\begin{equation}
\varDelta s=-\frac{x^*}{2\theta_x}\ ,
\end{equation}
as shown in Fig.~\ref{crabsext}. We use superscript $^*$ for the variables at the IP. Thus the associated transformation is
\begin{equation}
y^*\rightarrow y^*-p_y^*\varDelta s=y^*+\frac{p_y^*x^*}{2\theta_x}\ ,
\end{equation}
which is expressed as $\exp(:H^*:)$ with a Hamiltonian at the IP:
\begin{equation}
H^*=\frac{x^*p_y^{*2}}{4\theta_x}\ .
\end{equation}
If the phase advances between the IP and a sextupole (``{\it crab sextupole}") are:
\begin{equation}
\varDelta\psi_x=2\pi\ \ {\rm and}\ \ \varDelta\psi_y=2.5\pi\ ,\label{crabphase}
\end{equation}
then the variables at the IP $(x^*,p_y^*)$ can be expressed in those at the crab sextupole $(x,y)$:
\begin{equation}
x^* = \sqrt\frac{\beta_x^*}{\beta_x}x,\ \ p_y^*=\frac{y}{\sqrt{\beta_y^*\beta_y}}\ .
\end{equation}
Thus the Hamiltonian at the IP is equivalent to a Hamiltonian at the crab sextupole:
\begin{equation}
H=\frac{xy^2}{4\theta_x\beta_y^*\beta_y}\sqrt\frac{\beta_x^*}{\beta_x}\ ,
\end{equation}
which can be approximated by a Hamiltonian of a sextupole:
\begin{equation}
H_s=\frac{k_2}{6}\left(x^3-3xy^2\right)\ ,\label{Hs}
\end{equation}
by setting
\begin{equation}
k_2 = -\frac{1}{2\theta_x\beta_y^*\beta_y}\sqrt\frac{\beta_x^*}{\beta_x}\ .
\end{equation}
The $x^3$ term in Eq.~(\ref{Hs}) brings  only minor effect in the $x$-plane at the IP. The nonlinearity produced by the crab sextupole is absorbed by another crab sextupole located in the other side of the IP, which also has the sane phase relation as Eq. (\ref{crabphase}).  For the optics with $\beta_{x,y}^*=(1~{\rm m}, 2~{\rm mm})$, $k_2 = -0.85~{\rm m}^{-2}$. Then the crab sextupoles at (a,d) in Fig.~\ref{IR} becomes about 30\% {\it weaker} than the chromatic sextupoles ($\sim$2.5~m$^{-2}$) at (b,c).

\newpage
\section{Effect of Synchrotron Radiation in Quadrupoles on Dynamic Aperture\label{app:aSRQ}}
The synchrotron radiation power can be expressed as
\begin{equation}
P\propto \gamma^2B^2\ell\ ,
\end{equation}
where $\gamma$, $B$, $\ell$ are the Lorentz factor, transverse magnetic field, and the length of the field, respectively. Let us assume a particles is performing horizontal $\beta$-tron oscillation with amplitudes at the quadrupoles:
\begin{equation}
x_{\rm QF}=\varDelta x \cos(\psi_{\rm QF}), \qquad x_{\rm QD}=\varDelta x \sqrt{\frac{\beta_{x{\rm QD}}}{\beta_{x{\rm QF}}}} \cos(\psi_{\rm QD})\ ,\label{osci}
\end{equation}
through a periodic FODO lattice consisting of focusing/defocusing quadrupoles QF/QD. In Eq.~(\ref{osci}), $\psi_{\rm QF, QD}$ are the $\beta$-tron phases at the quadrupoles.  Then the average radiation powers in these quadrupoles are written as
\begin{eqnarray}
\left\langle P_{\rm QF}\right\rangle\propto&\gamma^2\left\langle(k_{\rm QF} B\rho/\ell_{\rm QF} x_{\rm QF})^2\right\rangle\ell_{\rm QF}=&\frac{\gamma^2}{2\ell_{\rm QF}}(k_{\rm QF} B\rho)^2\varDelta x^2\ , \label{PQF}\\
\left\langle P_{\rm QD}\right\rangle\propto&\gamma^2\left\langle(k_{\rm QD} B\rho/\ell_{\rm QD} x_{\rm QD})^2\right\rangle\ell_{\rm QD}=&\frac{\gamma^2}{2\ell_{\rm QD}}(k_{\rm QD} B\rho)^2\frac{\beta_{x{\rm QD}}}{\beta_{x{\rm QF}}}\varDelta x^2\ , \label{PQD}
\end{eqnarray}
where $\ell_{\rm QF,QD}$ and $\beta_{x{\rm QF,QD}}$ are the lengths and the horizontal $\beta$-functions of the quadrupoles, and $k_{\rm QF,QD}=B'_{\rm QF,QD}\ell_{\rm QF,QD}/B\rho$. The average over quadrupoles $\langle\rangle$ is taken by Eq.~(\ref{osci}). Similarly the average radiation loss in dipoles per a FODO cell is written as
\begin{equation}
\left\langle P_{\rm D}\right\rangle\propto\gamma^2 (B\rho\theta_c/\ell_c)^2\ell_c=\frac{\gamma^2}{\ell_c}(\theta_cB\rho)^2\ , \label{PD}
\end{equation}
where $\theta_c$ and $\ell_c$ are the bending angle and the length of a FODO cell. Here we assume the cell is fully filled by the dipoles and ignored the thickness of quadrupoles and other spaces.

\begin{figure}[h!]
   \centering
   \includegraphics[width=400pt]{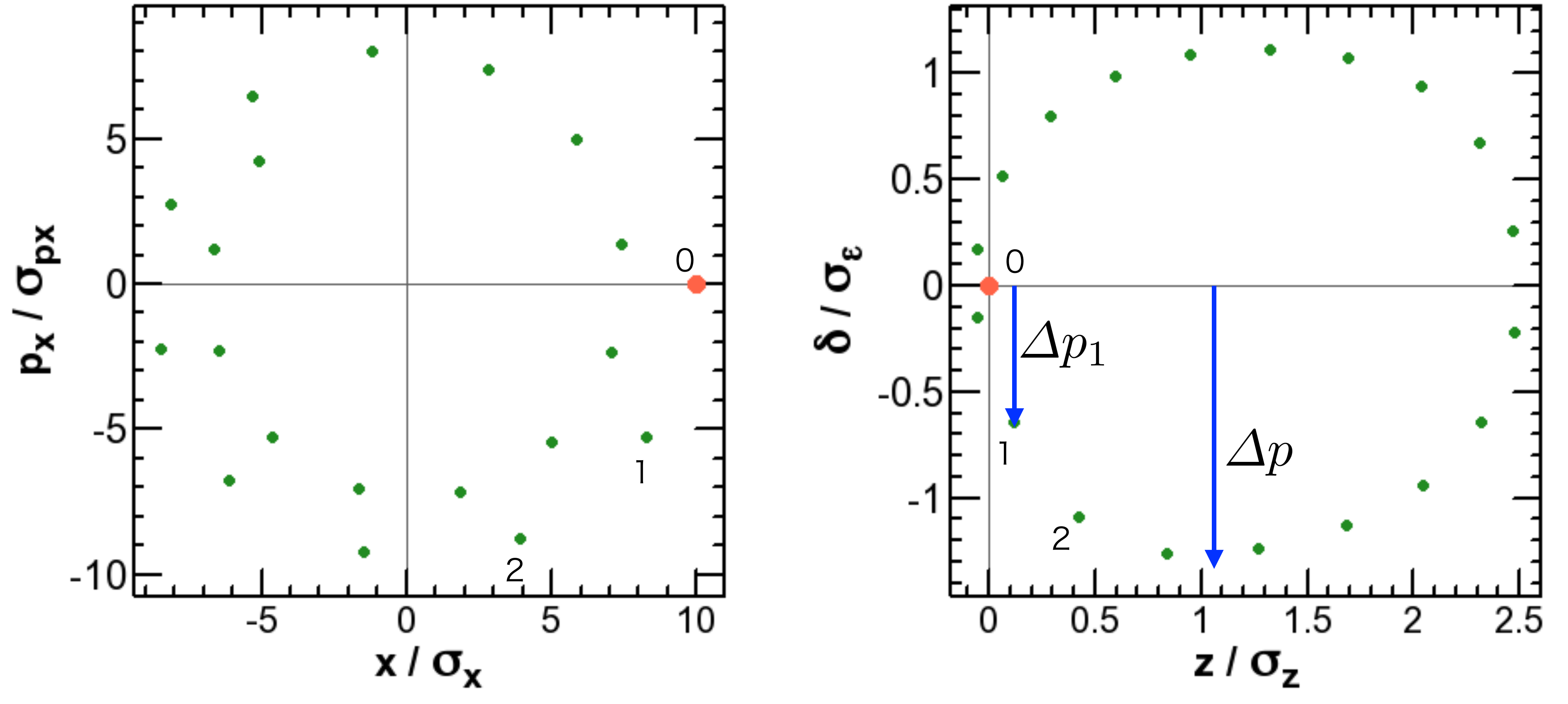}
   \caption{Poincar\'{e} plots in $x$-$p_x$ (left) and $z$-$\delta$ (right) planes for a particle starting at $x=10\sigma_x$, $p_x=y=p_y=z=\delta=0$, depicted by the red dots. The numbers, 0, 1, 2 are turns. The synchrotron radiation loss in the quadrupoles excites the synchrotron motion as shown in the right plot. The amount of the energy loss in the first turn $\varDelta p_1$, and the peak amplitude of the synchrotron motion $\varDelta p$ agrees with the estimation, Eq.~(\ref{dpamp}). \label{qrad}}
\end{figure}

Let us calculate the ratio of radiation losses in quadrupoles and dipoles per unit FODO cell, which contains one QF and QD each. Using Eqs. (\ref{PQF}--\ref{PD}),
\begin{equation}
\frac{\left\langle P_{\rm Q}\right\rangle}{\left\langle P_{\rm D}\right\rangle}=\frac{\left\langle P_{\rm QF}\right\rangle+\left\langle P_{\rm QD}\right\rangle}{\left\langle P_{\rm D}\right\rangle}=
\frac{\ell_c}{2\theta_c^2}\left(\beta_{x{\rm QF}}\frac{k_{\rm QF}^2}{\ell_{\rm QF}}+\beta_{x{\rm QD}}\frac{k_{\rm QD}^2}{\ell_{\rm QD}}\right) n^2\varepsilon_x\equiv R_{\rm Q} n^2\varepsilon_x\ , \label{RQ}
\end{equation}
where we have used $n\equiv\varDelta x/\sigma_x=\varDelta x/\sqrt{\beta_{x{\rm QF}}\varepsilon_x}$.
Then the relative radiation loss per ring $\varDelta P_{\rm Q}$ due to quadrupoles is obtained as
\begin{equation}
\varDelta p_1=\frac{\varDelta P_{\rm Q}}{E}=-\frac{U_0}{E} R_{\rm Q} n^2\varepsilon_x=-\frac{2\alpha_z}{J_z} R_{\rm Q} n^2\varepsilon_x , \label{loss1}
\end{equation}
where $U_0$, $\alpha_z$, and $J_z$  are the radiation loss and synchrotron damping rate per revolution, and the longitudinal damping partition number. It is possible to express the parameters $k_{\rm QF,QD}$ and $\beta_{x{\rm QF,QD}}$ in Eq.~(\ref{RQ}) in terms of $\ell_c$, once the $\beta$-tron phase advances of the FODO are chosen. In the case of the $90^\circ/90^\circ$ as in this design, they are written by a thin-lens approximation as
\begin{equation}
k_{\rm QF}=-k_{\rm QD}=\frac{2\sqrt2}{\ell_c},\qquad\beta_{x{\rm QF}}=\left(1+\frac{1}{\sqrt2}\right)\ell_c,\qquad \beta_{x{\rm QD}}=\left(1-\frac{1}{\sqrt2}\right)\ell_c,
\end{equation}
then we get
\begin{equation}
R_{\rm Q}=\frac{2\sqrt2}{\theta_c^2}\left(\frac{\sqrt2+1}{\ell_{\rm QF}}+\frac{\sqrt2-1}{\ell_{\rm QD}}\right)\ ,
\end{equation}
which is independent of $\ell_c$.
 As the horizontal betatron motion induces the energy loss of Eq.~(\ref{loss1}), the associated synchrotron motion peaks at the $1/4\nu_s$-th turn with an amplitude
 \begin{equation}
 \varDelta p=\frac{\varDelta p_1}{2\pi\nu_s}\exp(-\alpha_z/4\nu_s)=\frac{\alpha_z}{\pi\nu_sJ_z} R_{\rm Q} n^2\varepsilon_x\exp(-\alpha_z/4\nu_s)\ ,\label{dpamp}
 \end{equation}
where $\nu_s=|\nu_z|$ is the absolute value of the synchrotron tune. We have neglected the effect of transverse damping in Eq.~(\ref{dpamp}). If we plug in the numbers for this design at 175 GeV listed in Table~\ref{srparam}, we obtain
\begin{equation}
\varDelta p_1 = -0.58\sigma_\varepsilon{\rm\ \ \ and\ \ \ }\varDelta p = -1.38\sigma_\varepsilon \label{dpnum}
\end{equation}
with $n=10$, which agrees with the tracking simulation shown in Fig.~\ref{qrad}.

\begin{table}[h!]
   \centering
  \caption{Parameters related to the synchrotron radiation loss in the quadrupoles. \label{srparam}}
\vspace{6pt}
   \begin{tabular}{|l c|l c|}
  \hline
$\alpha_z$ & 0.0427 &
$\nu_s$ & 0.0546\\
$J_z$ & 1.98 &
$\varepsilon_x$ [nm] & 1.26 \\
$\theta_c$ [mrad] & 4.27 & $\sigma_\varepsilon$ [\%]& 0.141\\
$\ell_{\rm QF}$ [m] & 3.5 &
$\ell_{\rm QD}$ [m] & 1.8\\
\hline
\end{tabular}
\end{table}

\begin{figure}[h!]
   \centering
   \includegraphics[width=450pt]{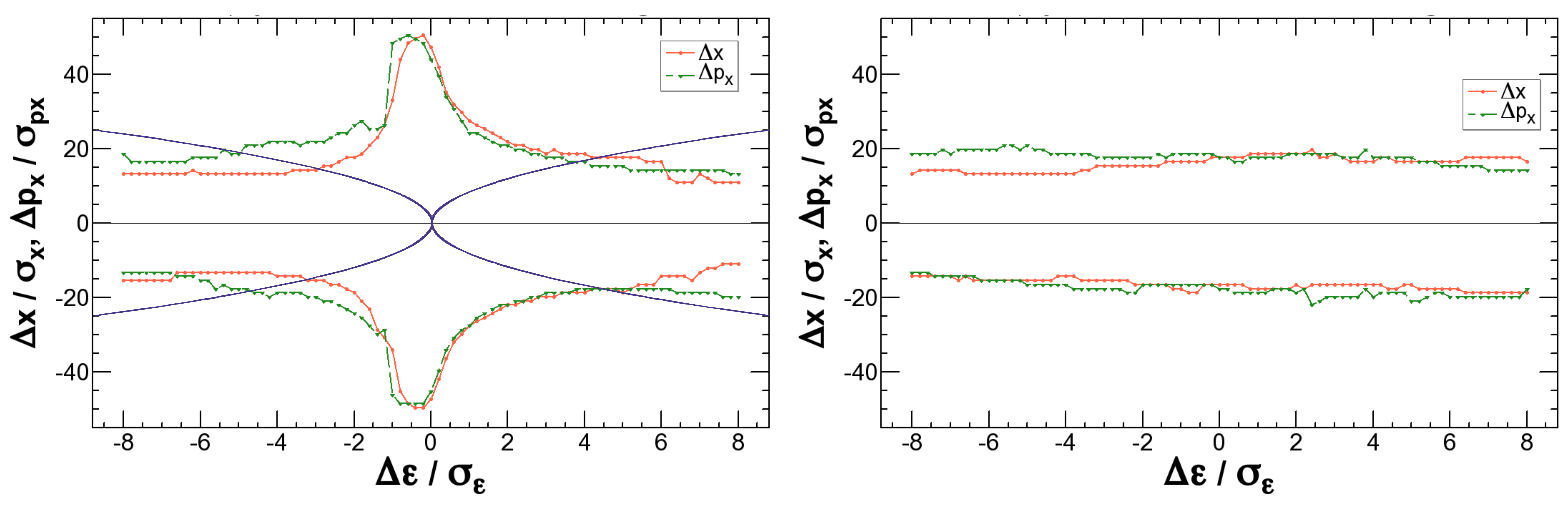}
   \caption{The dynamic aperture in the $z$-$x$ plane without(left)/with(right) synchrotron radiation in the quadrupoles at 175~GeV. The parabolas on the left show the amplitude of the synchrotron motion given by Eq.~(\ref{dpamp}). For a value of $n=\varDelta x/\sigma_x$, if an unstable region exists within these parabolas, then the motion larger than $n$ becomes unstable. \label{srdyn}}
\end{figure}

 The synchrotron motion caused by the synchrotron radiation in the quadrupoles has a significant impact on the dynamic aperture at 175 GeV. This kind of effect has been already noticed at LEP2~\cite{ref:LEPQSR}. The dynamic aperture without radiation in the quadrupoles has a sharp peak for the on-momentum particles due to the non-interleaved sextupoles. The synchrotron motion associated with the radiation in the quadrupoles, however, destroys such a peak, as no particles with large horizontal amplitude stay on-momentum. This situation is explained by Fig.~\ref{srdyn}. 
 
\section{Effect of radiation fluctuation on dynamic aperture\label{fluct}}
The effect of the radiation fluctuation on DA is studied by adding a random component on the radiation loss at each magnet. Since the number of photons per revolution is $\sim$21,000 at $t\overline t$, we assumed that a random number corresponding to the expected value of photon energy spread is enough for the simulation. We did not simulate the full quantized photon emission, including the randomness in the emission timing and exact photon energy spectrum, to save computation time. 

\begin{figure}[h!]
   \centering
   \includegraphics[width=300pt]{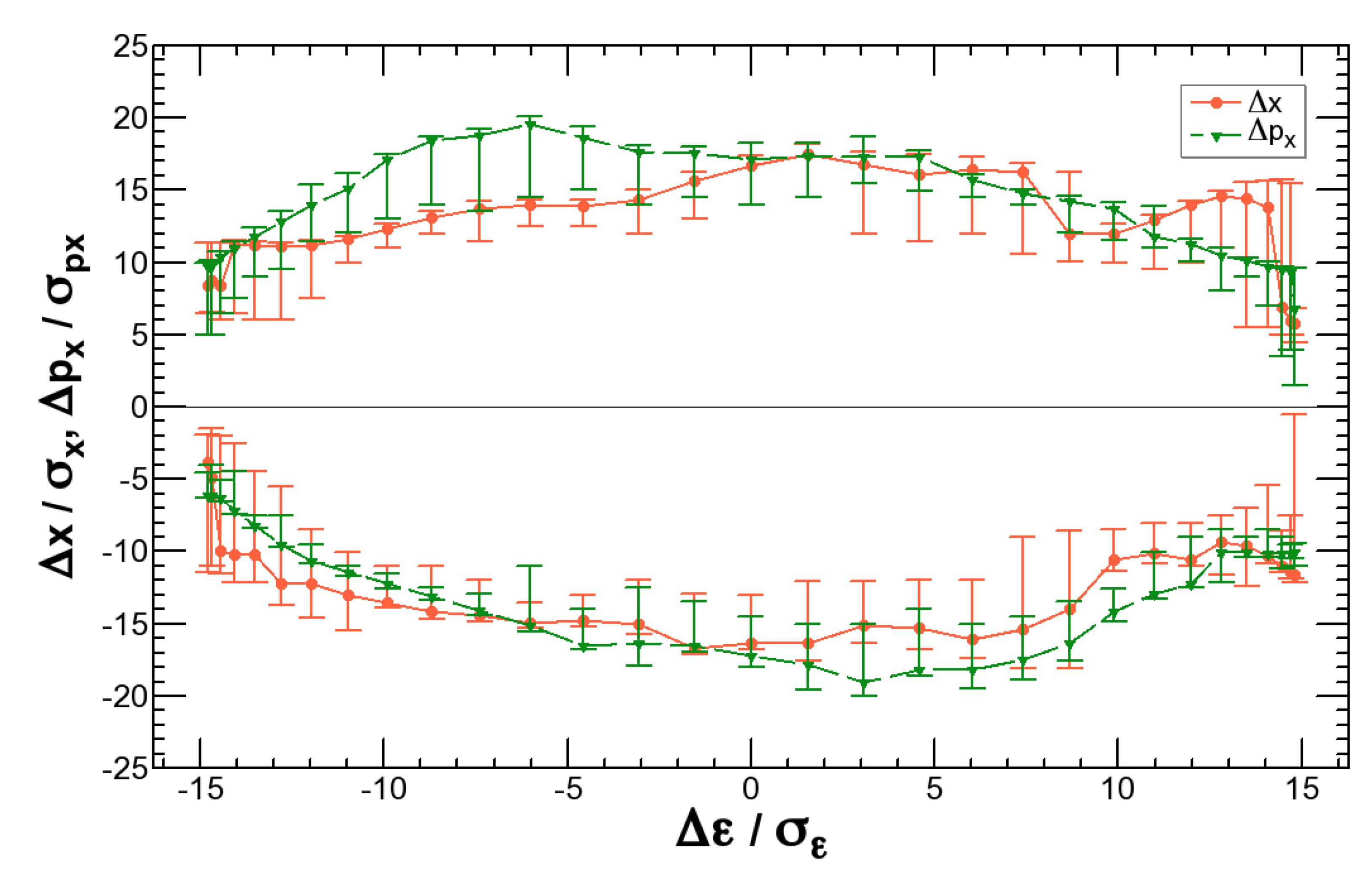}
   \caption{The dynamic aperture in the $z$-$x$ plane at 175~GeV, $\beta^*_{x,y} =$ (1~m, 2~mm), with synchrotron radiation fluctuation. The lines correspond to the aperture where 75\% of 100 samples were stable for 50 turns. The error bars indicate the range of stability between 50\% and 100\% for 100 samples. The case without fluctuation is plotted in Fig.~\ref{DA}(a).\label{DAf}}
\end{figure}

Figure~\ref{DAf} shows the DA estimated at ${t\overline t}$ for 100 samples for the fluctuation in a similar way as described in Sec.~\ref{DAsec}. The aperture corresponding to 50\% stable samples is more or less similar to that without fluctuation. This is expected, as 50\% of the particles will be lost if they started at a rigid physical aperture. The 100\% stable aperture still satisfies the requirements for the DA in this design.

On the other hand, the width between 50\% and 100\% stability reaches to $\sim$5$\sigma_x$ at nearly all energy deviations,  which is larger than na\"{\i}vely expected in the case of a physical aperture with a simple diffusion/damping model.
 We did not analyze the problem in detail, but a possible explanation is that there could be a number of unstable orbits in the phase space around the grid points surveyed by the tracking. The optimization of DA may try to shift such unstable orbits away from the surveyed orbits, but they can still exist in the near region. Thus particles can hit such unstable orbit by the radiation fluctuation, and once they hit the unstable orbit, the growth rate can easily exceed the radiation damping. The density of such unstable orbits should naturally shrink as the amplitude becomes small. 

\begin{acknowledgments}
The authors thank D. Schulte for providing information on FCC-hh. We also thank R.~Calaga, C. Cook, S. Fartoukh, P. Janot, E.~Jensen, R. Kersevan, H.~Koiso, A.~Milanese, P.~Raimondi, J.~Seeman, D. Shwartz, G. Stupakov, R. Tomas for useful discussions and suggestions.
\end{acknowledgments}


\bibliography{FCCee_Optics}

\end{document}